\documentclass[a4paper,twoside]{article}

\usepackage[dvips]{graphicx}
\usepackage{color}
\usepackage[utf8]{inputenc}
\usepackage[T2A,TS1,T1]{fontenc}
\usepackage[english,russian]{babel}
\usepackage{amsmath,amsthm,amssymb}
\usepackage{euscript}
\usepackage{calc}

\PassOptionsToPackage{lowtilde}{url}
\RequirePackage[%
    colorlinks=true,
    pdfstartview=FitH,
    linkcolor=blue,
    citecolor=blue,
    filecolor=blue,
    urlcolor=blue,
    linktoc=page,
    pdftitle={Asymptotic bounds of depth for a reversible circuit consisting of NOT, CNOT and 2-CNOT gates},
    pdfauthor={Dmitry V. Zakablukov},
    pdfkeywords={reversible logic, circuit depth, computations with memory},
    pdfsubject={The paper discusses the asymptotic depth of a reversible circuit consisting of NOT, CNOT and 2-CNOT gates}]{hyperref}

\newcommand{\compose}{\mathop{*}\limits}
\newcommand{\ZZ}{\mathbb Z}
\newcommand{\frS}{\mathfrak S}
\newcommand{\vv}{\mathbf}

\title{Оценка глубины обратимых схем из функциональных элементов NOT, CNOT и 2-CNOT}
\author{Д.\,В.~Закаблуков\protect\footnote{%
    {\it Закаблуков Дмитрий Владимирович} --- асп. каф. информационной безопасности ф-та информатики и систем управления
    МГТУ им. Н.\,Э.~Баумана,
    e-mail: \texttt{\href{mailto:dmitriy.zakablukov@gmail.com}{dmitriy.zakablukov@gmail.com}}.}
}

\theoremstyle{plain}
\newtheorem{theorem}{Теорема}

\theoremstyle{definition}

\begin{document}

\maketitle

\begin{abstract}
Рассматривается вопрос об асимптотической глубине обратимых схем, состоящих из функциональных элементов NOT, CNOT и 2-CNOT.
Вводится функция Шеннона $D(n, q)$ глубины обратимой схемы, реализующей какое-либо отображение $f\colon \ZZ_2^n \to \ZZ_2^n$,
как функция от $n$ и количества дополнительных входов схемы $q$.
Доказывается, что при реализации отображения $f$, задающего четную подстановку на множестве $\ZZ_2^n$,
обратимой схемой, не использующей дополнительные входы, верно соотношение $D(n, 0) \gtrsim 2^n / (3\log_2 n)$.
Устанавливается также, что при использовании $q_0 \sim 2^n$ дополнительных входов для реализации произвольного отображения
$f\colon \ZZ_2^n \to \ZZ_2^n$ в обратимой схеме верно соотношение $D(n, q_0) \lesssim 3n$.
\end{abstract}

\textbf{Ключевые слова}: обратимые схемы, глубина схемы, вычисления с памятью.

\textbf{1. Введение.}
В дискретной математике нередко возникает задача оценивания асимптотической сложности того или иного преобразования.
Теория схемной сложности берет свое начало с работы K.~Шеннона~\cite{shannon}. В ней в качестве меры сложности
булевой функции предлагается рассматривать сложность минимальной контактной схемы, реализующей эту функцию.
О.\,Б. Лупановым~\cite{lupanov_method} установлена асимптотика сложности $L(n) \sim \rho 2^n \mathop / n$
булевой функции от $n$ переменных в произвольном конечном полном базисе элементов с произвольными положительными весами,
где $\rho$ обозначает минимальный приведенный вес элементов базиса.

Вопрос о вычислениях с ограниченной памятью был рассмотрен Н.\,А.~Карповой в работе~\cite{karpova},
где доказано, что в базисе классических функциональных элементов, реализующих все \mbox{$p$-местные} булевы функции,
асимптотическая оценка функции Шеннона сложности схемы с тремя и более регистрами памяти
зависит от значения $p$, но не изменяется при увеличении количества используемых регистров памяти.
Также было показано, что существует булева функция, которая не может быть реализована в маломестных базисах
с использованием менее двух регистров памяти.

О.\,Б. Лупановым в работе~\cite{lupanov_delay} рассмотрены схемы из функциональных элементов с задержками. Было доказано,
что в регулярном базисе функциональных элементов любая булева функция может быть реализована схемой,
имеющей задержку $T(n) \sim \tau n$, где $\tau$~--- минимум приведенных задержек всех элементов базиса, при сохранении
асимптотически оптимальной сложности. Однако не рассматривался вопрос зависимости $T(n)$
от количества используемых регистров памяти. Хотя задержка и глубина схемы в некоторых работах определяются по-разному%
~\cite{khrapchenko}, в исследуемой далее модели обратимой схемы их, по мнению автора, можно отождествить.

В настоящей работе рассматриваются схемы, состоящие из обратимых функциональных элементов NOT (инвертор), 1-CNOT (контролируемый
инвертор, CNOT) и 2-CNOT (дважды контролируемый инвертор, элемент Тоффоли).
Будут использоваться формальные определения этих элементов и состоящих из них схем из работы~\cite{my_fast_alg}.
В~\cite{shende,my_research} доказано, что для любой четной подстановки $h \in A(\ZZ_2^n)$ существует задающая ее
обратимая схема с $n$ входами, состоящая из элементов NOT, CNOT и 2-CNOT.

В работе~\cite{vinokurov} рассматривались обратимые схемы без дополнительных входов (дополнительной памяти),
состоящие из обобщенных элементов Тоффоли $k$-CNOT; была установлена нижняя оценка функции Шеннона сложности таких схем
$L(n) > \frac{n2^n}{\log_2 n + n - 1}$. В~\cite{shende} также рассматривались обратимые схемы без дополнительных входов,
но уже в базисе NOT, CNOT и 2-CNOT; были доказаны нижняя оценка функции Шеннона сложности таких схем
$L(n) = \Omega\left(\frac{n2^n}{\log_2 n}\right)$ и верхняя оценка $L(n) \leqslant O(n2^n)$.
В~\cite{maslov_tech} была улучшена верхняя оценка: $L(n) \lesssim 5n2^n$.
Однако схемы с дополнительными входами в данных работах не рассматривались.

В работе~\cite{my_complexity} было доказано, что для функции Шеннона сложности обратимых схем без дополнительных входов,
состоящих из элементов NOT, CNOT и 2-CNOT, верно соотношение $L(n) \asymp \frac{n2^n}{\log_2 n}$.
Также было показано, что использование дополнительной памяти в таких схемах почти всегда позволяет снизить сложность схемы.

Автору не удалось найти какие-либо опубликованные результаты об оценке функции Шеннона глубины обратимых схем,
состоящих из элементов NOT, CNOT и 2-CNOT. Тем не менее в работе~\cite{abdessaied} было экспериментально показано,
что использование $O(n)$ дополнительных входов позволяет значительно снизить глубину таких схем.

В настоящей работе рассматривается множество $F(n,q)$ всех отображений $\ZZ_2^n \to \ZZ_2^n$, которые могут быть реализованы
обратимой схемой, состоящей из элементов NOT, CNOT и 2-CNOT (далее просто обратимая схема), с $(n+q)$ входами.
Оценивается глубина обратимой схемы, реализующей отображение $f \in F(n,q)$ 
с использованием $q$ дополнительных входов. Вводится функция Шеннона $D(n, q)$ глубины обратимой схемы
как функция от $n$ и количества дополнительных входов схемы $q$. Показывается, что, как и в случае сложности
обратимой схемы~\cite{my_complexity}, глубина обратимой схемы существенно зависит от количества дополнительных входов
(регистров памяти, см.~\cite{karpova}).

При помощи мощностного метода Риордана--Шеннона доказывается нижняя оценка глубины обратимой схемы
$D(n, q) \geqslant (2^n(n-2) - n\log_2(n+q)) \mathop / (3(n + q) \log_2 (n + q))$.
Описывается аналогичный методу О.\,Б. Лупанова~\cite{lupanov_delay} подход к синтезу обратимой схемы, для которого глубина
синтезированной схемы $D(n, q_0) \lesssim 3n$ при использовании $q_0 \sim 2^n$ дополнительных входов.

\textbf{2. Основные понятия.}
Определение обратимых функциональных элементов было впервые введено Т.~Тоффоли~\cite{toffoli}.
Обратимые функциональные элементы NOT и $k$-CNOT, а также синтез схем из этих элементов были рассмотрены, к примеру,
в работе~\cite{maslov_thesis}.

Будем пользоваться следующим формальным определением функциональных элементов NOT и $k$-CNOT~\cite{my_fast_alg}.
Через $N_j^n$ обозначается функциональный элемент NOT (инвертор) с $n$ входами,
задающий преобразование $\ZZ_2^n \to \ZZ_2^n$ вида
\begin{equation}
    N_j^n(\langle x_1, \ldots, x_n \rangle) = \langle x_1, \ldots, x_j \oplus 1, \ldots, x_n \rangle \; .
    \label{eq1}
\end{equation}
Через $C_{i_1,\ldots,i_k;j}^n = C_{I;j}^n$, $j \notin I$, обозначается функциональный элемент $k$-CNOT с $n$ входами
(контролируемый инвертор, обобщенный элемент Тоффоли с $k$ контролирующими входами),
задающий преобразование $\ZZ_2^n \to \ZZ_2^n$ вида
\begin{equation}
    C_{i_1,\ldots,i_k;j}^n(\langle x_1, \ldots, x_n \rangle) =
        \langle x_1, \ldots, x_j \oplus x_{i_1} \wedge \ldots \wedge x_{i_k}, \ldots, x_n \rangle \; .
    \label{eq2}
\end{equation}
Далее будем опускать верхний индекс $n$, если его значение ясно из контекста.
Обозначим через $\Omega_n^2$ множество всех функциональных элементов NOT, CNOT и 2-CNOT с $n$ входами.

Схема из функциональных элементов классически определяется как ориентированный граф без циклов с помеченными ребрами и вершинами.
В обратимых схемах, состоящих из элементов множества $\Omega_n^2$, запрещено ветвление и произвольное подключение
входов и выходов функциональных элементов. В ориентированном графе, описывающем такую обратимую схему $\frS$,
все вершины, соответствующие функциональным элементам, имеют ровно $n$ занумерованных входов и выходов.
Эти вершины нумеруются от 1 до $l$, при этом $i$-й выход $m$-й вершины, $m < l$,
соединяется только с $i$-м входом $(m+1)$-й вершины. Входами обратимой схемы являются входы первой вершины,
а выходами~--- выходы $l$-й вершины.
Соединение функциональных элементов друг с другом будем также называть композицией элементов.

Всем $i$-м входам и выходам вершин графа приписывается символ $r_i$ из некоторого множества $R = \{\,r_1, \ldots, r_n\,\}$.
Каждый символ $r_i$ можно интерпретировать как имя регистра памяти (номер ячейки памяти), хранящего текущий результат
работы схемы. Из формул~\eqref{eq1} и~\eqref{eq2} следует, что в один момент времени
(один такт работы схемы) может быть инвертировано значение не более чем в одном регистре памяти.
В этом заключается существенное отличие обратимых схем от схем из классических функциональных элементов,
рассмотренных О.\,Б. Лупановым в своих работах.

Среди основных характеристик обратимой схемы можно выделить сложность и глубину схемы. Пусть обратимая схема $\frS$
с $n$ входами представляет собой композицию $l$ элементов из множества $\Omega_n^2$:
$\frS = \compose_{j=1}^l {E_j(t_j, I_j)}$, где $t_j$ и $I_j$~--- контролируемый выход и множество контролирующих входов
элемента $E_j$ соответственно.
Сложность $L(\frS)$ обратимой схемы $\frS$~--- количество элементов в схеме $l$.
Классически глубина схемы из функциональных элементов определяется как длина максимального пути на графе,
описывающем данную схему, между какими-либо входными и выходными вершинами. В рассматриваемой модели обратимой схемы граф,
описывающий такую схему, представляет собой просто одну цепочку последовательно соединенных вершин. Поэтому, если использовать
классическое определение глубины схемы, получится, что в нашем случае глубина обратимой схемы равна ее сложности.

Для того чтобы не менять модель обратимой схемы, введем следующее определение глубины обратимой схемы.
Будем считать, что обратимая схема $\frS = \compose_{j=1}^l {E_j(t_j, I_j)}$ имеет глубину $D(\frS) = 1$,
если для любых двух ее функциональных элементов $E_1(t_1,I_1)$ и $E_2(t_2,I_2)$ выполняется равенство
$$
    \left(\{\,t_1\,\} \cup I_1 \right) \cap \left(\{\,t_2\,\} \cup I_2 \right) = \varnothing \; .
$$
Также будем считать, что обратимая схема $\frS$ имеет глубину $D(\frS) \leqslant d$, если ее можно разбить на
$d$ непересекающихся подсхем, каждая из которых имеет глубину 1:
\begin{equation}
    \frS = \bigsqcup_{i=1}^d{{\frS'}_i} \; , \text{ }D({\frS'}_i) = 1 \; .
    \label{eq3}
\end{equation}
Тогда можно ввести следующее определение: глубина $D(\frS)$ обратимой схемы $\frS$~--- минимально возможное количество $d$
непересекающихся подсхем глубины 1 в разбиении схемы $\frS$ по формуле~\eqref{eq3}.
Используя это определение, можно вывести простое соотношение, связывающее сложность и глубину обратимой схемы $\frS$,
имеющей $n$ входов:
\begin{equation}
    \frac{L(\frS)}{n} \leqslant D(\frS) \leqslant L(\frS) \; .
    \label{eq4}
\end{equation}

\begin{figure}
    \centering
    \includegraphics[scale=1]{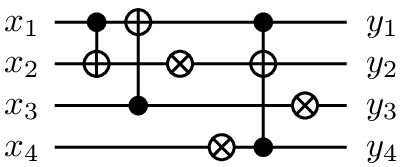}
    \caption
    {
        Обратимая схема $\frS = C_{1;2} * C_{3;1} * N_2 * N_4 * C_{1,4;2} * N_3$
        со сложностью $L(\frS) = 6$ и глубиной $D(\frS) = 3$
    }\label{pic1}
\end{figure}

На рис.~\ref{pic1} показан пример обратимой схемы со сложностью 6 и глубиной 3. На данном и на всех последующих рисунках
элементы $k$-CNOT обозначаются следующим образом: контролирующие входы обозначаются символом {\large $\bullet$},
контролируемый выход~--- символом {\small $\boldsymbol{\oplus}$}.
Инвертируемый выход элемента NOT обозначается символом {\small $\boldsymbol{\otimes}$}.
Входы схемы/элементов, если не оговорено иначе, находятся слева, выходы~--- справа.
Входы и выходы пронумерованы сверху вниз начиная с 1.
Элементы в схеме соединяются без ветвлений входов и выходов, $i$-й выход $j$-го элемента соединяется
с $i$-м входом $(j+1)$-го элемента.
На входы обратимой схемы подаются значения 0 и 1, затем последовательно, слева направо, каждый из элементов инвертирует либо
не инвертирует значение на одном (и только одном) из своих выходов в зависимости от значений на своих входах
(см. формулы~\eqref{eq1} и~\eqref{eq2}).

\textbf{3. Глубина обратимой схемы.}
Введем следующие отображения:

1) \textit{расширяющее} отображение $\phi_{n,n+k}\colon \ZZ_2^n \to \ZZ_2^{n+k}$ вида
$$
    \phi_{n,n+k}( \langle x_1, \ldots, x_n \rangle ) = \langle x_1, \ldots, x_n, 0, \ldots, 0 \rangle \; ;
$$

2) \textit{редуцирующее} отображение $\psi_{n+k,n}^\pi\colon \ZZ_2^{n+k} \to \ZZ_2^n$ вида
$$
    \psi_{n+k,n}^\pi( \langle x_1, \ldots, x_{n+k} \rangle ) =
    \langle x_{\pi(1)}, \ldots, x_{\pi(n)} \rangle \; ,
$$
где $\pi$~--- некоторая подстановка на множестве $\ZZ_{n+k}$.

Известно, что обратимая схема с $n \geqslant 4$ входами задает некоторую четную подстановку
на множестве $\ZZ_2^n$~\cite{shende,my_research}.
В тоже время данная схема может реализовывать не более $A_n^m$ (количество размещений из $n$ по $m$ без повторений)
различных булевых отображений $\ZZ_2^n \to \ZZ_2^m$, где $m \leqslant n$, с использованием или без использования
дополнительных входов.
Введем формальное определение обратимой схемы, реализующей некоторое отображение $f\colon \ZZ_2^n \to \ZZ_2^m$
с использованием дополнительных входов.

\textbf{Определение 1.}
Обратимая схема $\frS_g$ с $(n+q)$ входами, задающая преобразование $g\colon \ZZ_2^{n+q} \to \ZZ_2^{n+q}$,
\textit{реализует} отображение $f\colon \ZZ_2^n \to \ZZ_2^m$ c использованием $q \geqslant 0$ дополнительных входов
(дополнительной памяти), если существует такая подстановка $\pi \in S(\ZZ_{n+q})$, что
$$
	\psi_{n+q,m}^\pi(g( \phi_{n,n+q}(\vv x))) = f(\vv x), \text{ где } \vv x \in \ZZ_2^n, f(\vv x) \in \ZZ_2^m \; .
$$

Выражения <<реализует отображение>> и <<задает отображение>> имеют различное значение: если обратимая схема $\frS_g$
задает отображение $f$, то $g(\vv x) = f(\vv x)$.
Будем говорить, что схема $\frS_g$ реализует отображение $f$ \textit{без использования дополнительной памяти},
если она имеет ровно $n$ входов.
Очевидно, что при $m > n$ не существует обратимой схемы, реализующей отображение $f$ без использования дополнительной памяти.

Обозначим через $P_2(n,n)$ множество всех булевых отображений $\ZZ_2^n \to \ZZ_2^n$.
Обозначим через $F(n,q) \subseteq P_2(n,n)$ множество всех отображений $\ZZ_2^n \to \ZZ_2^n$, которые могут быть реализованы
обратимой схемой с $(n+q)$ входами.
Множество подстановок из $S(\ZZ_2^n)$, задаваемых всеми элементами множества $\Omega_n^2$,
генерирует знакопеременную $A(\ZZ_2^n)$ и симметрическую $S(\ZZ_2^n)$ группы подстановок при $n > 3$ и $n \leqslant 3$
соответственно~\cite{shende,my_research}.
Отсюда следует, что $F(n,0)$ совпадает с множеством преобразований, задаваемых всеми подстановками из $A(\ZZ_2^n)$
и $S(\ZZ_2^n)$ при $n > 3$ и $n \leqslant 3$ соответственно.
С другой стороны, несложно показать, что при $q \geqslant n$ верно равенство $F(n,q) = P_2(n,n)$.

Обозначим через $L(f,q)$ и $D(f,q)$ минимальную сложность и глубину обратимой схемы,
состоящей из функциональных элементов множества $\Omega_{n+q}^2$ и реализующей некоторое отображение $f \in F(n,q)$
с использованием $q$ дополнительных входов.
Определим функции Шеннона $L(n,q)$ и $D(n,q)$ для сложности и глубины обратимой схемы:
$$
    L(n,q) = \max_{f \in F(n,q)} {L(f,q)} \; , \; \;  D(n,q) = \max_{f \in F(n,q)} {D(f,q)} \; .
$$

Сформулируем основные результаты работы. Доказательство приведенных ниже теорем будет дано в следующих пунктах.
Будем использовать следующие обозначения для асимптотического неравенства, эквивалентности
и эквивалентности с точностью до порядка двух функций от $n$: $f(n) \gtrsim g(n)$, $f(n) \sim g(n)$ и $f(n) \asymp g(n)$.

\begin{theorem}{Теорема 1 (нижняя оценка сложности обратимой схемы).}\label{thm1}
    Для любого $n > 0$ верно неравенство
    $$
        L(n,q) \geqslant \frac{2^n(n - 2) - n \log_2(n+q)}{3\log_2(n+q)} \; .
    $$
\end{theorem}

\begin{theorem}{Следствие 1.}
    Для любого $n > 0$ верно неравенство
    $$
        D(n,q) \geqslant \frac{2^n(n - 2) - n \log_2(n+q)}{3(n+q)\log_2(n+q)} \; .
    $$
\end{theorem}
Доказательство следует из теоремы~\ref{thm1} и соотношения~\eqref{eq4}.

\begin{theorem}{Следствие 2.}
    Для обратимой схемы $\frS$ без дополнительных входов верна следующая нижняя оценка глубины:
    $$
        D(n,0) \gtrsim \frac{2^n}{3\log_2 n} \; .
    $$
\end{theorem}

\begin{theorem}{Теорема 2 (верхняя оценка глубины обратимой схемы).}\label{thm2}
	Верны следующие оценки:
    \begin{align*}
        D(n,q_1) &\lesssim 3n \text{ \,при\, } q_1 \sim 2^n, \text{ }
            L(\frS) \sim 2^{n+1} \; , \\
        D(n,q_2) &\lesssim 2n \text{ \,при\, } q_2 \sim \phi(n)2^n, \text{ }
            L(\frS) \sim \phi(n)2^{n+1}  \; ,
    \end{align*}
    где $\phi(n) < n$~--- сколь угодно медленно растущая функция от $n$.
\end{theorem}

\begin{theorem}{Утверждение.}
    Использование дополнительной памяти в обратимых схемах, состоящих из функциональных элементов множества $\Omega_n^2$,
    почти всегда позволяет существенно снизить глубину обратимой схемы, в отличие от схем,
    состоящих из классических необратимых функциональных элементов~\textnormal{\cite{lupanov_delay}}.
\end{theorem}
Доказательство следует из теорем~\ref{thm1} и~\ref{thm2}.

Мы не оцениваем глубину обратимых схем, реализующих отображения $\ZZ_2^n \to \ZZ_2^m$ при $m \neq n$.
Тем не менее для таких схем могут быть получены аналогичные оценки глубины путем корректной подстановки параметра
$m$ в доказательство теорем~\ref{thm1}, 2.

Полученные верхние и нижние оценки глубины обратимой схемы достаточно неточны: они фактически несопоставимы.
Так, нижняя оценка глубины при количестве дополнительных входов $q_2 \sim \phi(n)2^n$ из теоремы~\ref{thm2}
вырождается в тривиальную $D(n,q_2) \geqslant 0$,
в то время как верхняя оценка при данном значении количества дополнительных входов линейна.

К сожалению, вопрос получения эквивалентных с точностью до порядка верхних и нижних оценок для $D(n,q)$ до
сих пор остается открытым. Автор надеется, что результаты данной работы станут первым шагом в данном направлении.

\textbf{4. Нижняя оценка сложности обратимых схем.}
Перейдем к доказательству первой теоремы.

{\bf Доказательство теоремы~\ref{thm1}.}
    Докажем при помощи мощностного метода Риордана--Шеннона, что для любого $n > 0$ верно неравенство
    $$
        L(n,q) \geqslant \frac{2^n(n - 2) - n \log_2(n+q)}{3\log_2(n+q)} \; .
    $$

    Пусть $r = |\Omega_n^2|$.
    Обозначим через $\EuScript C^*(n,s) = r^s$ и $\EuScript C(n,s)$ количество всех обратимых схем,
    которые состоят из функциональных элементов множества $\Omega_n^2$
	и сложность которых равна $s$ и не превышает $s$ соответственно.
    Тогда
    \begin{gather*}
        r = |\Omega_n^2| = \sum_{k=0}^2{(n-k)\binom{n}{k}} = \frac{n^3 - n^2 + 2n}{2} \; , \\
        \frac{n^2(n-1)}{2} + 1 < r \leqslant \frac{n^3}{2} \text{ \,при\, } n \geqslant 2 \; , \\
        \EuScript C(n,s) = \sum_{i=0}^s {\EuScript C^*(n,i)} = \frac{r^{s+1} - 1}{r-1}
            \leqslant \left( \frac{n^3}{2} \right)^{s+1} \cdot \frac{2}{n^2(n-1)} \; , \\
        \EuScript C(n,s) \leqslant \left( \frac{n^3}{2} \right)^s \cdot \left(1 + \frac{1}{n-1}\right)
            \text{ \,при\, } n \geqslant 2 \; .
    \end{gather*}

    Как было сказано ранее, каждой обратимой схеме с $(n+q)$ входами
    соответствует не более $A_{n+q}^n$ различных булевых отображений $\ZZ_2^n \to \ZZ_2^n$.
    Пусть $s = L(n,q)$, тогда верно следующее неравенство:
    $$
       \EuScript C(n+q,s) \cdot A_{n+q}^n \geqslant |F(n,q)| \; .
    $$
    Поскольку $|F(n,q)| \geqslant |A(\ZZ_2^n)| = (2^n)! \mathop / 2$ и $A_{n+q}^n \leqslant (n+q)^n$, то
    $$
        \left( \frac{(n+q)^3}{2} \right)^s \cdot \left(1 + \frac{1}{n+q-1}\right) \cdot (n+q)^n \geqslant (2^n)! \mathop / 2 \; .
    $$
    Несложно убедиться, что при $n > 0$ верно неравенство $(2^n)! > (2^n \mathop / e)^{2^n}$. Следовательно,
    \begin{gather*}
        s(3\log_2(n+q) - 1) + \log_2 \left(1 + \frac{1}{n+q-1}\right) + n \log_2(n+q) \geqslant 2^n(n - \log_2 e) \; , \\
        s \geqslant \frac{2^n(n - 2) - n \log_2(n+q)}{3\log_2(n+q)} \; .
    \end{gather*}
    \noindent Из этого неравенства следует утверждение теоремы, поскольку в наших обозначениях $s = L(n,q)$.
\qed

В работе~\cite{my_complexity} была сделана попытка поднять нижнюю оценку сложности обратимых схем за счет
свойства эквивалентности
некоторых схем с точки зрения задаваемых ими преобразований. Для этой цели была выдвинута следующая гипотеза о структуре
обратимых схем из функциональных элементов множества $\Omega_n^2$.

\textbf{Гипотеза.}
    \textit{Почти каждая обратимая схема, состоящая из функциональных элементов} NOT\textit{,} CNOT \textit{и}
	2-CNOT \textit{и имеющая $n \to \infty$ входов,
    может быть представлена в виде композиции подсхем сложности $k = o(n)$} (\textit{кроме последней, у которой сложность $L \leq k$})\textit{,
    таких, что в каждой подсхеме все элементы являются попарно коммутирующими.
    Количество обратимых схем, для которых это неверно, пренебрежимо мало.}

К сожалению, в доказательстве этой гипотезы в работе~\cite{my_complexity} была допущена ошибка:
несложно показать, что количество всех обратимых схем сложности выше $n$, не соответствующих утверждению гипотезы,
не является пренебрежимо малым по отношению к общему количеству схем данной сложности.


\textbf{5. Верхняя оценка глубины обратимых схем без дополнительных входов.}
В работе~\cite{my_complexity} предложен алгоритм синтеза обратимой схемы $\frS$ без дополнительных входов, задающей подстановку
$h \in A(\ZZ_2^n)$; доказано, что сложность синтезированной схемы удовлетворяет соотношению
\begin{equation}
    L(\frS) \lesssim 52n 2^n / \log_2 n \; .
    \label{eq5}
\end{equation}
Очевидно, что $D(\frS) \leqslant L(\frS)$, поэтому $D(n,0) \lesssim 52n 2^n / \log_2 n$.
Однако константу 52 в данной оценке можно уменьшить.

Алгоритм синтеза из работы~\cite{my_complexity} задает произведение $L \sim \log_2 n$ независимых транспозиций одним
элементом $k$-CNOT и множеством элементов NOT и CNOT с помощью действия сопряжением.
Для этого строится матрица из векторов, соответствующих этим транспозициям. В матрице обнуляются некоторые
столбцы путем сложения по модулю 2 с совпадающими с ними столбцами (не более $2n$ элементов CNOT),
а в конце работы алгоритма почти все столбцы матрицы делаются единичными (не более $2n$ элементов NOT)%
\footnote[2]{Коэффициент 2 возникает за счет действия сопряжением.}.
Очевидно, что обнулять столбцы матрицы можно с логарифмической глубиной (глубина не более $2 \log_2 n$),
а элементы NOT можно применять с константной глубиной (глубина не превышает 2).

Если аккуратно заменить в доказательстве оценки~\eqref{eq5} из работы~\cite{my_complexity} величину $4n$,
соответствующую сложности описанных выше шагов алгоритма синтеза обратимой схемы,
на величину $2(\log_2 n + 1)$, то можно получить следующую верхнюю оценку для $D(n,0)$:
$$
    D(n,0) \lesssim 36n 2^n / \log_2 n \; .
$$

Однако остается открытым вопрос получения эквивалентных с точностью до порядка нижней и верхней оценок для функции $D(n,0)$.


\textbf{6. Верхняя оценка глубины обратимых схем с дополнительными входами.}
О.\,Б. Лупановым~\cite{lupanov_delay} был предложен асимптотически наилучший метод синтеза схем из функциональных элементов
с задержками, реализующих булевы функции, в регулярном базисе. Было доказано, что для булевой функции от $n$ переменных
и в случае равных единичных задержек всех элементов базиса задержка схемы эквивалентна $n$.
Применим аналогичный подход для получения верхней оценки глубины обратимых схем, состоящих из функциональных элементов
множества $\Omega_{n+q}^2$ и реализующих заданное отображение $f \in F(n,q)$.

\begin{figure}
    \centering
    \includegraphics[scale=1]{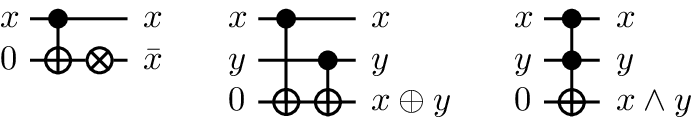}
    \caption
    {
        Выражение функциональных элементов базиса $\{\,\neg, \oplus, \wedge\,\}$ через
        композицию элементов NOT, CNOT и 2-CNOT
    }\label{pic2}
\end{figure}

Базис $\{\,\neg, \oplus, \wedge\,\}$ является функционально полным, поэтому с его помощью можно
реализовать любое отображение $f \in F(n,q)$. Выразим каждый элемент этого базиса
через композицию функциональных элементов NOT, CNOT и 2-CNOT (рис.~\ref{pic2}). Видно, что каждый
элемент реализуется со сложность и глубиной не выше 2, при этом требуется максимум один дополнительный вход.

\begin{figure}
    \centering
    \includegraphics[scale=1]{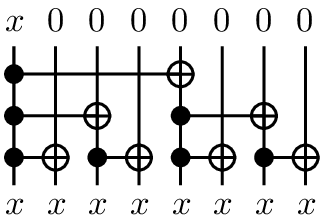}
    \caption
    {
        Копирование значения с одного входа на дополнительные входы с логарифмической глубиной
        (входы схемы сверху, выходы~--- снизу)
    }\label{pic3}
\end{figure}

Отметим также, что если значение с одного входа в дальнейшем должно участвовать в $k$ операциях, то для уменьшения
глубины схемы производится копирование этого значения на дополнительные входы, а затем эти дополнительные входы используются в
$k$ операциях независимо друг от друга. В итоге можно получить подсхему с глубиной не $k$,
а $(\lceil \log_2 k \rceil + 1)$ (рис.~\ref{pic3}).

Докажем следующую лемму о глубине обратимой схемы, реализующей все конъюнкции $n$ переменных вида
$x_1^{a_1} \wedge \ldots \wedge x_n^{a_n}$, $a_i \in \ZZ_2$.
\begin{theorem}{Лемма.}
Все конъюнкции $n$ переменных вида $x_1^{a_1} \wedge \ldots \wedge x_n^{a_n}$, $a_i \in \ZZ_2$,
можно реализовать обратимой схемой $\frS_n$, состоящей из функциональных элементов множества $\Omega_n^2$,
имеющей глубину $D(\frS_n) \sim n$ и использующей $q(\frS_n) \sim 3 \cdot 2^n$ дополнительных входов.
При этом сложность такой схемы $L(\frS_n) \sim 3 \cdot 2^n$.
\end{theorem}
{\bf Доказательство.}
    Вначале реализуем все инверсии $\bar x_i$, $1 \leqslant i \leqslant n$. Сделать это можно с глубиной $D_1 = 2$
    при использовании $L_1 = 2n$ элементов NOT и CNOT и $q_1 = n$ дополнительных входов.

    Искомую обратимую схему $\frS_n$ будем строить следующим образом: при помощи схем $\frS_{\lceil n/2 \rceil}$
    и $\frS_{\lfloor n/2 \rfloor}$ реализуем все конъюнкции $\lceil n/2 \rceil$ первых и $\lfloor n/2 \rfloor$ последних
    переменных. Затем реализуем конъюнкции выходов этих двух схем каждого с каждым.
    Любой выход будет участвовать не более чем в $2 \cdot 2^{n/2}$ конъюнкциях, поэтому получение искомых конъюнкций
    можно реализовать с глубиной не более чем $(2 + n/2)$, сложностью не более чем $3 \cdot 2^n$ и с использованием не более
    чем $3 \cdot 2^n$ дополнительных входов.

    Таким образом, получаем следущие соотношения:
    \begin{gather*}
        D(\frS_n) \sim \frac{n}{2} + D(\frS_{n/2}) \sim n \; , \\
        L(\frS_n) \sim 3 \cdot 2^n + 2L(\frS_{n/2}) \sim 3 \cdot 2^n \; , \\
        q(\frS_n) \sim 3 \cdot 2^n + 2q(\frS_{n/2}) \sim 3 \cdot 2^n \; .
    \end{gather*}
\qed

Теперь перейдем непосредственно к доказательству теоремы~\ref{thm2}. Основное отличие метода синтеза,
описываемого в этом доказательстве, от стандартного метода О.\,Б. Лупанова заключается в следующем:
в обратимых схемах запрещено ветвление входов и выходов, поэтому для получения требуемых оценок для функции $D(n,q)$
активно используются подсхемы по копированию значений с промежуточных выходов на дополнительные входы
с логарифмической глубиной (см. рис.~\ref{pic3}).
Также подсчитывается количество используемых дополнительных входов и получаемая
при этом сложность схемы.

{\bf Доказательство теоремы~\ref{thm2}.}
    Докажем, что для произвольного отображения $f \in F(n,q)$ верны следующие соотношения:
    \begin{equation}
        D(f,q_1) \lesssim 3n \text{ при } q_1 \sim 2^n, \text{ }
            L(\frS) \sim 2^{n+1} \; ,
        \label{eq6}
    \end{equation}
    \begin{equation}
        D(f,q_2) \lesssim 2n \text{ при } q_2 \sim \phi(n)2^n, \text{ }
            L(\frS) \sim \phi(n)2^{n+1} \; ,
        \label{eq7}
    \end{equation}
    где $\phi(n) < n$~--- сколь угодно медленно растущая функция от $n$.

    Булево отображение $f\colon \ZZ_2^n \to \ZZ_2^n$ можно представить следующим образом:
    \begin{equation}
        f(\vv x) = \bigoplus_{a_{k+1}, \ldots, a_n \in \ZZ_2} {x_{k+1}^{a_{k+1}} \wedge \ldots \wedge x_n^{a_n}}
        \wedge f(\langle x_1, \ldots, x_k, a_{k+1}, \ldots, a_n \rangle) \; .
        \label{eq8}
    \end{equation}

    Каждое из $2^{n-k}$ отображений
    $f_i(\langle x_1, \ldots, x_k \rangle) = f(\langle x_1, \ldots, x_k, a_{k+1}, \ldots, a_n \rangle)$,
    где
    $$
        \sum_{j=1}^{n-k} {a_{k+j} \cdot 2^{j-1}} = i \; ,
    $$
    является отображением $\ZZ_2^k \to \ZZ_2^n$. Его можно представить в виде системы $n$ координатных булевых функций
    $f_{i,j}(\vv x)$, $\vv x \in \ZZ_2^k$, $1 \leqslant j \leqslant n$.

    Воспользуемся следующим аналогом совершенной дизъюнктивной нормальной формы для булевой функции:
    \begin{equation}
        f_{i,j}(\vv x) = \bigoplus_{
            \substack{\boldsymbol \sigma \in \ZZ_2^k \\f_{i,j}(\boldsymbol \sigma) = 1}}
            x_1^{\sigma_1} \wedge \ldots \wedge x_k^{\sigma_k} \; .
        \label{eq9}
    \end{equation}

    Разбив все $2^k$ конъюнкций вида $x_1^{\sigma_1} \wedge \ldots \wedge x_k^{\sigma_k}$ на фиксированные группы,
    в каждой из которых не более $s$ конъюнкций, получим $p = \lceil 2^k / s \rceil$ групп.
    Используя конъюнкции одной группы, по формуле~\eqref{eq9} можно получить не более $2^s$ булевых функций.
    Обозначим множество булевых функций, реализуемых при помощи конъюнкций $i$-й по счету группы, через $G_i$,
    $1 \leqslant i \leqslant p$, тогда $|G_i| \leqslant 2^s$.
    Теперь мы можем переписать равенство~\eqref{eq9} в следующем виде:
    \begin{equation}
        f_{i,j}(\vv x) = \bigoplus_{
            \substack{t=1 \ldots p\\ g_{j_t} \in G_t\\ 1 \leqslant j_t \leqslant |G_t|}} g_{j_t}(\vv x) \; .
        \label{eq10}
    \end{equation}

    \textit{Замечание.} Все булевы функции множества $G_i$ можно реализовать по тому же принципу, что и все конъюнкции в
    лемме (разбиение множества входов пополам):
    глубина полученной подсхемы $D \sim s$, сложность $L \sim 3 \cdot 2^s$, количество дополнительных входов $q \sim 2^{s+1}$.

\begin{figure}
        \centering
        \includegraphics[scale=1]{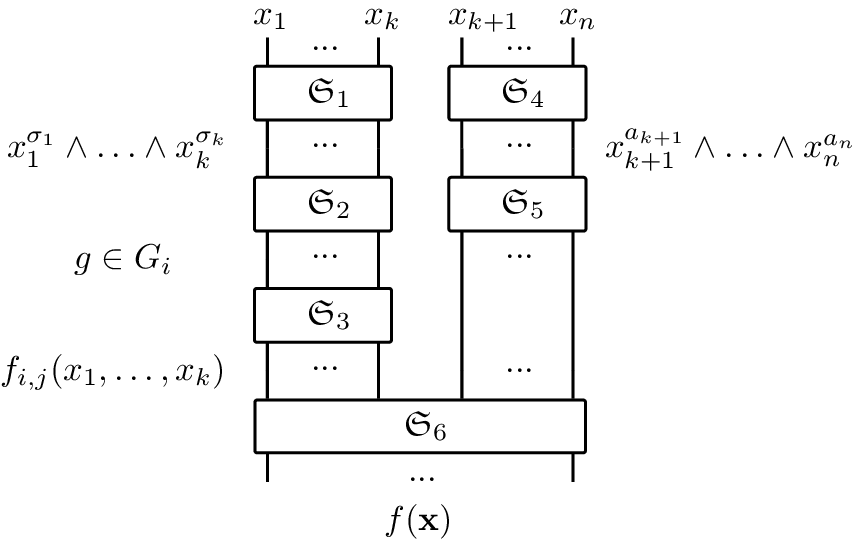}
        \caption
        {
            Структура обратимой схемы $\frS$, реализующей отображение~\eqref{eq8} (входы схемы сверху, выходы~--- снизу)
        }\label{pic4}
    \end{figure}

    Таким образом, искомая обратимая схема $\frS$, реализующая отображение $f$, состоит из следующих обратимых подсхем
    (рис.~\ref{pic4}).
	
	1) Подсхема $\frS_1$, реализующая все конъюнкции первых $k$ переменных $x_i$, согласно лемме,
	с глубиной $D_1 \sim k$, сложностью $L_1 \sim 3 \cdot 2^k$ и $q_1 \sim 3 \cdot 2^k$ дополнительными входами.
	
	2) Подсхема $\frS_2$, реализующая все булевы функции $g \in G_i$ для всех $i \in \ZZ_p$
	по формуле~\eqref{eq9} с глубиной $D_2 \sim s$, сложностью $L_2 \sim 3p2^s$
	и $q_2 \sim p2^{s+1}$ дополнительными входами (см. замечание, касающееся реализации всех булевых функций множества $G_i$).
	
	3) Подсхема $\frS_3$, реализующая все $n2^{n-k}$ координатных функций $f_{i,j}(\vv x)$,
	$i \in \ZZ_{2^{n-k}}$, $j \in \ZZ_n$, по формуле~\eqref{eq10}.
	Особенностью данной подсхемы является то, что некоторая булева функция $g \in G_t$ может использоваться больше
	одного раза. Максимальное количество использования функции $g$ не превосходит $n2^{n-k}$.
	Следовательно, сперва нам необходимо скопировать значения с выходов подсхемы $\frS_2$
	для всех таких булевых функций.
	Это можно сделать с глубиной $(n-k +\log_2 n)$, используя не более $pn2^{n-k}$ функциональных элементов
	и $pn2^{n-k}$ дополнительных входов (см. рис.~\ref{pic3}).
	Затем производится сложение по модулю 2 полученных выходов с глубиной $\log_2 p$,
	сложностью $(p-1)n2^{n-k}$ и без дополнительных входов.
	Таким образом, подсхема $\frS_3$ имеет глубину $D_3 \sim n-k + \log_2 p$,
	сложность $L_3 \sim (2p-1)n 2^{n-k}$ и $q_3 \sim pn 2^{n-k}$ дополнительных входов.

    4) Подсхема $\frS_4$, реализующая все конъюнкции последних $(n-k)$ переменных $x_i$, согласно лемме,
	с глубиной $D_4 \sim (n-k)$, сложностью $L_4 \sim 3 \cdot 2^{n-k}$ и $q_4 \sim 3 \cdot 2^{n-k}$
	дополнительными входами.

	5) Подсхема $\frS_5$, необходимая для копирования $(n-1)$ раз значения каждого выхода подсхемы $\frS_4$.
	Это можно сделать с глубиной $D_5 \sim \log_2 n$, сложностью $L_5 = (n-1) \cdot 2^{n-k}$
	и $q_5 = (n-1)2^{n-k}$ дополнительными входами.

	6) Подсхема $\frS_6$, реализующая булево отображение $f$ по формуле~\eqref{eq8}.
	Структура данной подсхемы следующая: все $n2^{n-k}$ координатных функций $f_{i,j}(\vv x)$ группируются
	по $2^{n-k}$ функций (всего $n$ групп, соответствующих $n$ выходам отображения $f$).
	Функции одной группы объединяются по две.
	В каждой паре функций производится конъюнкция соответствующих выходов подсхем $\frS_3$ и $\frS_5$
	при помощи двух элементов 2-CNOT. При этом для каждой пары функций используется один дополнительный вход
	для хранения промежуточного результата.
	Таким образом, данный этап требует глубины 2, $n2^{n-k}$ элементов 2-CNOT и $n2^{n-k-1}$ дополнительных входов.
	Затем в каждой из $n$ групп полученных значений происходит суммирование по подулю 2 при помощи элементов CNOT
	с логарифмической глубиной.
	Следовательно, этот этап требует глубины $(n-k-1)$, элементов CNOT в количестве $n(2^{n-k-1} - 1)$
	и не использует дополнительные входы,
	так как можно обойтись уже существующими выходами для суммирования по модулю 2.
	
	В итоге получаем подсхему $\frS_6$ с глубиной $D_6 \sim (n-k)$, сложностью $L_6 \sim 3n 2^{n-k-1}$
	и $q_6 \sim n 2^{n-k-1}$ дополнительными входами.

    Отметим, что подсхемы $\frS_1$--$\frS_3$ и $\frS_4$--$\frS_5$ могут работать параллельно,
    поскольку они работают с непересекающимися подмножествами множества входов $x_1, \ldots, x_n$
    обратимой схемы $\frS$ (см. рис.~\ref{pic4}).

    Будем искать параметры $k$ и $s$, удовлетворяющие следующим условиям:
    $$
        \left\{
            \begin{array}{lr}
                k + s = n \; , & \\
                1 \leqslant k < n \; , & \\
                1 \leqslant s < n \; , & \\
                2^k \mathop / s \geqslant \psi(n) \;, & \text{ где $\psi(n)$~--- некоторая растущая функция.}
            \end{array}
        \right.
    $$
    В этом случае $p = \lceil 2^k \mathop / s \rceil \sim 2^k \mathop / s$.
    
    Суммируя глубины, сложности и количество дополнительных входов всех подсхем $\frS_1$--$\frS_6$,
    получаем следующие оценки для характеристик обратимой схемы $\frS$.
	
    Глубина:
    \begin{gather*}
        D(\frS) \sim \max(k + s + n - k + \log_2 p \;;\; n-k + \log_n) + n - k \; , \\
        D(\frS) \sim 2n + s \; .
    \end{gather*}

    Сложность:
    \begin{gather*}
        L(\frS) \sim 3 \cdot 2^k + 3p2^s + (2p-1)n2^{n-k} + 3 \cdot 2^{n-k} + n2^{n-k} + 3n2^{n-k-1} \; , \\
        L(\frS) \sim 3 \cdot \frac{2^n}{2^s} + \frac{3 \cdot 2^n}{s} + \frac{n2^{n+1}}{s} \sim \frac{n2^{n+1}}{s} \; .
    \end{gather*}

    Количество используемых дополнительных входов:
    \begin{gather*}
        q(\frS) \sim 3 \cdot 2^k + p2^{s+1} + pn2^{n-k} + 3 \cdot 2^{n-k} + n2^{n-k} + n2^{n-k-1} \; , \\
        q(\frS) \sim 3 \cdot \frac{2^n}{2^s} + \frac{2^{n+1}}{s} + \frac{n2^n}{s} \sim \frac{n2^n}{s} \; .
    \end{gather*}

    Мы построили обратимую схему $\frS$ для произвольного отображения $f \in F(n,q)$, откуда следует, что
    $D(n,q) \leqslant D(\frS)$.
    
    Оценка~\eqref{eq6} достигается при $k = \lceil n \mathop / \phi(n)\rceil$, $s = n - \lceil n \mathop / \phi(n)\rceil$,
    где $\phi(n) \leqslant n \mathop / (\log_2 n + \log_2 \psi(n))$ и $\psi(n)$~--- любые сколь угодно медленно растущие функции.
    
    Оценка~\eqref{eq7} достигается при $k = n - \lceil n \mathop / \phi(n)\rceil$, $s = \lceil n \mathop / \phi(n)\rceil$,
    где $\phi(n) < n$~--- сколь угодно медленно растущая функция.
\qed

Остается открытым вопрос получения эквивалентных с точностью до порядка нижней и верхней оценок для функции $D(n,q)$ в
случае $q \to \infty$.


\textbf{7. Заключение.}
В работе рассмотрен вопрос о глубине обратимых схем, состоящих из функциональных элементов NOT, CNOT и 2-CNOT.
Изучена функция Шеннона $D(n,q)$ глубины обратимой схемы, реализующей какое-либо отображение $\ZZ_2^n \to \ZZ_2^n$ из
множества $F(n,q)$, как функции от $n$ и количества дополнительных входов схемы $q$.
Доказаны некоторые нижние и верхние асимптотические оценки функции $D(n,q)$ для обратимых схем с дополнительными входами и без.
Показано, что использование дополнительной памяти в таких обратимых схемах почти всегда позволяет снизить глубину
схемы, в отличие от схем, состоящих из классических необратимых функциональных элементов.

При решении задачи синтеза обратимой схемы, реализующей какое-либо отображение, приходится искать компромисс
между сложностью синтезированной схемы, ее временем работы (глубина схемы) и количеством используемой
дополнительной памяти (дополнительных входов в схеме).
Направлением дальнейших исследований является более детальное изучение зависимости этих величин друг от друга.


\end{document}